\documentclass[%
 reprint,
superscriptaddress,
 amsmath,amssymb,
 aps,
pra,
floatfix,
]{revtex4-1}

\usepackage{graphicx}
\usepackage{dcolumn}
\usepackage{bm}
\usepackage{color}
\usepackage[linktocpage,colorlinks=true,linkcolor=blue,citecolor=blue,urlcolor=blue,breaklinks=true]{hyperref}   
\usepackage{MnSymbol}

\begin{document}
\bibliographystyle{unsrt}

\title{Collective Criticality through Adaptive Investor-Market Feedback}

\author{Jiazhen Liu}
\email{liujiazhen@fudan.edu.cn}
\affiliation{%
Research Institute of Intelligent Complex Systems, Fudan University, Shanghai 200433, China.}%
\author{Zhongpu Qiu}
\affiliation{%
Shanghai AI Laboratory, Shanghai 200433, China}
\email{slszqzp@outlook.com}
\author{Jin Guo}
\email{jin.guo@xjtlu.edu.cn}
\affiliation{%
Department of Finance, International Business School Suzhou (IBSS), Xi’an Jiaotong-Liverpool University
(XJTLU), Suzhou, China
}%

\begin{abstract}
How adaptive feedback between investor behavior and market returns gives rise to collective criticality remains an open question. Here, we show how reciprocal adaptation between investor strategies and market returns generates market-mediated collective self-organization, driving a critical transition in investor sentiment without imposing direct alignment interactions among investors. A minimal stochastic framework reveals an emergent effective collective coupling and identifies the dynamical condition for the instability of the neutral collective state. Crucially, a single collective critical mode links the reorganization of investor sentiment to the scaling of macroscopic market responses and critical slowing down, unifying the static and dynamical signatures of the transition. Numerical simulations and empirical financial data exhibit behavior consistent with these predictions. Our results provide a dynamical interpretation of financial market criticality as a macroscopic manifestation of collective behavioral reorganization of investors.
\end{abstract}

\maketitle
Financial markets provide a prominent example of a complex adaptive system in which individual decisions and collective outcomes continuously influence one another \cite{lux1999scaling,savit1999adaptive,bardoscia2017pathways}. While investor sentiment and herding have long been associated with market instability \cite{sornette2023non,eguiluz2000transmission,guo2024institutional}, a fundamental question is whether critical market responses can be understood as macroscopic manifestations of a critical reorganization of collective investor behavior. Addressing this question requires moving beyond the amplification of individual behavioral biases to identify how adaptive investor-market interactions organize these biases into a persistent collective state, and how the resulting transition is transmitted to observable market dynamics.

Statistical-physics approaches to financial markets have demonstrated how interacting and adaptive agents can generate complex market fluctuations, collective organization, and phase transitions \cite{nagel2025macroscopic,cividino2023multiasset,fabiani2024task,challet2000statistical}. These studies have established a foundation for understanding the emergence of macroscopic financial phenomena from microscopic investor behavior and interactions \cite{sato2025strict,sato2023inferring,bucci2019crossover,toth2011anomalous,wand2024estimating,farmer2005predictive,patzelt2018universal}. In particular, Ising-type models have provided a statistical-physics framework for describing investor herding and collective financial bubbles through interacting behavioral states \cite{cividino2023multiasset,SornetteZhou2006}. Complementary behavioral-finance models have shown how extrapolative expectations and adaptive investor demand can amplify price dynamics through endogenous feedback \cite{barberis2015x,barberis2018extrapolation}. Yet it remains unclear how reciprocal adaptation between investor strategies and market returns can mediate collective self-organization and how the resulting critical dynamics jointly govern macroscopic market responses and collective relaxation. Addressing these questions requires identifying the dynamical mechanism that links market-mediated behavioral organization to observable static and dynamical signatures of criticality.

A minimal mechanism lies in the coevolution of investor strategies and market returns \cite{scholl2021market,kanazawa2018derivation,kanazawa2018kinetic,bouchaud1998langevin,baule2024feedback}, that is, individual strategy adjustments collectively shape market returns, while the resulting market response feeds back into subsequent decisions across the investor population. This reciprocal adaptation makes the market an endogenous mediator of effective collective coupling, through which initially weak behavioral deviations can become self-reinforcing and develop into a persistent collective state. Crucially, such self-organization does not arise from feedback amplification alone, but from a dynamical instability when market-mediated reinforcement overcomes individual behavioral stabilization. Identifying this instability and its consequences for observable market dynamics requires an adaptive dynamic framework that treats investor behavior and market returns as mutually coupled dynamical variables.

In this Letter, we show how reciprocal adaptation between investor strategies and market returns generates market-mediated collective self-organization and drives a critical transition in investor sentiment. Starting from a minimal nonequilibrium stochastic model, we identify an effective collective coupling emerging from the investor–market feedback loop and derive the dynamical condition under which this coupling destabilizes the neutral collective state. Crucially, the resulting collective critical mode links the emergence of persistent directional sentiment to the scaling of macroscopic market responses and the slowing down of collective relaxation, providing a unified dynamical explanation for these distinct critical signatures. Numerical simulations and empirical financial data further test the predicted collective transition and its observable consequences, establishing a quantitative connection between microscopic behavioral adaptation and macroscopic market criticality.
\begin{figure}
  \includegraphics[width=1\linewidth]{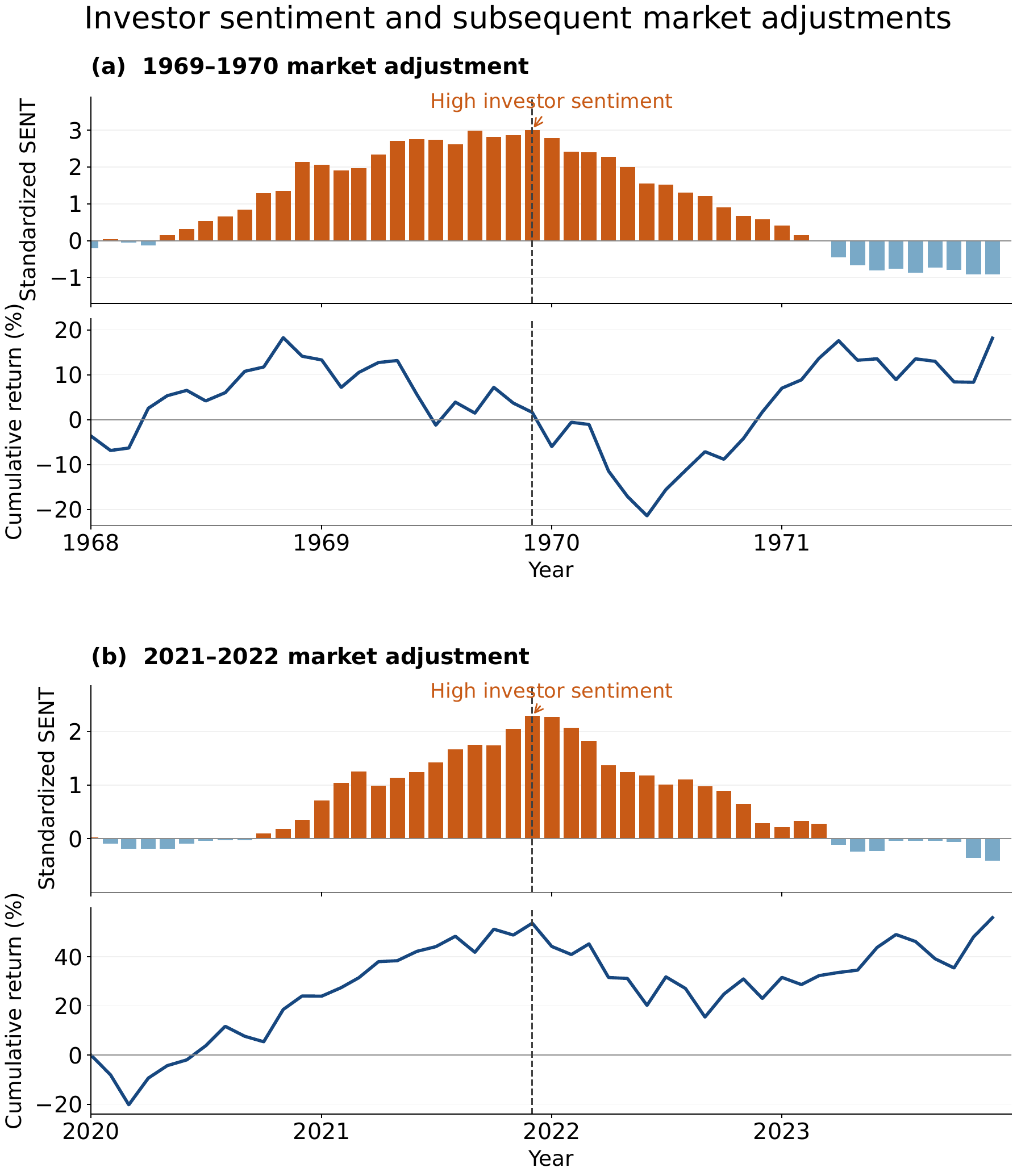}
\caption{Investor sentiment and market dynamics during historical U.S. bear markets. (a) The 1968--1970 bear market and (b) the 2022 bear market. The upper panels show the standardized Baker--Wurgler investor sentiment index (SENT), while the lower panels display cumulative U.S. stock market returns. Vertical dashed lines mark the peaks of investor sentiment within the respective pre-adjustment periods. Both episodes illustrate pronounced positive sentiment biases followed by substantial market declines, motivating an investigation of the collective dynamics arising from adaptive investor--market feedback.}
\label{fig:historical_cases}
  \label{fig1}
\end{figure}

{\it Empirical Observations.} To illustrate the interplay between collective investor sentiment and market dynamics, we examine two historical U.S. bear markets, the 1968--1970 and 2022 episodes. We characterize aggregate investor sentiment using the Baker--Wurgler sentiment index (SENT) \cite{fama1993common,baker2006investor} and track market performance through cumulative stock market returns (see Supplemental Material~\cite{LiuSupplementalMaterial}, Sec.~2.1). As shown in Fig.~\ref{fig1}, both episodes exhibit pronounced positive sentiment biases, accompanied by subsequent market declines. In particular, elevated sentiment is followed by substantial market adjustments, suggesting a close temporal association between collective investor behavior and market dynamics. These observations raise a fundamental question: Can adaptive feedback between investor sentiment and market returns drive collective state transitions and critical market responses?

{\it Modeling framework.}
To understand how individual behavioral biases can develop into collective market instability, we construct a minimal stochastic model of adaptive investor–market dynamics. Each investor is characterized by a time-dependent strategy bias \(w\in[-1,1]\), with \(w=-1\) and \(w=1\) representing extreme bearish and bullish tendencies, respectively, and \(w=0\) denoting neutrality. The collective sentiment \(W\) measures the directional organization of these individual strategies, distinguishing a neutral collective state (\(W\simeq0\)) from a collectively biased state (\(|W|>0\)). 

Crucially, this organization is not determined by investor behavior alone. Individual strategies adapt to market returns, while their collective bias feeds back into subsequent market movements. This reciprocal dependence creates a self-reinforcing mechanism through which initially weak behavioral biases may become collectively amplified. We therefore model investor strategies and market returns as mutually coupled stochastic dynamics, seeking to determine when this feedback destabilizes the neutral collective state and how the resulting collective organization affects macroscopic market responses.
\begin{figure}
  \includegraphics[width=1\linewidth]{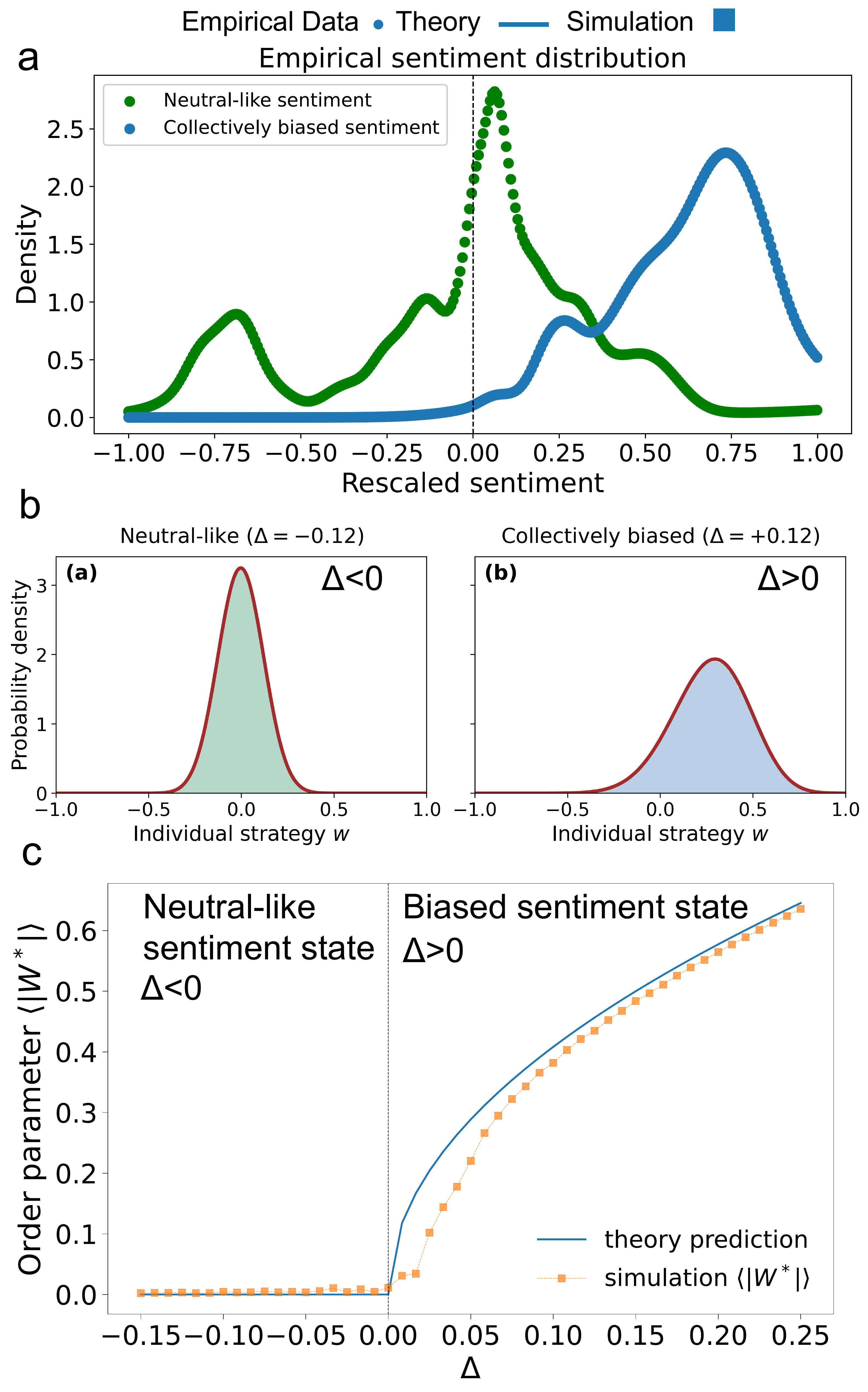}
  \caption{Collective reorganization of investor sentiment and its critical transition.
(a) Empirical distributions of investor sentiment during historically documented episodes (see Supplemental Material~\cite{LiuSupplementalMaterial}, Sec.~2.1) of market instability and speculative activity (blue, biased-sentiment) and non-event reference periods (green, neutral-like sentiment).
(b) Stationary distributions of collective sentiment predicted by the theory (solid curves) and reproduced by microscopic simulations (bar plots), for $\Delta<0$ (green) and $\Delta>0$ (blue).
(c) Continuous emergence of collective organization across the critical point. The order parameter $|W^*|$ obtained from simulations (symbols) agrees with the theoretical prediction (solid curve). The dashed line marks $\Delta=0$.
}
  \label{fig2}
\end{figure}

\begin{figure}
  \includegraphics[width=1\linewidth]{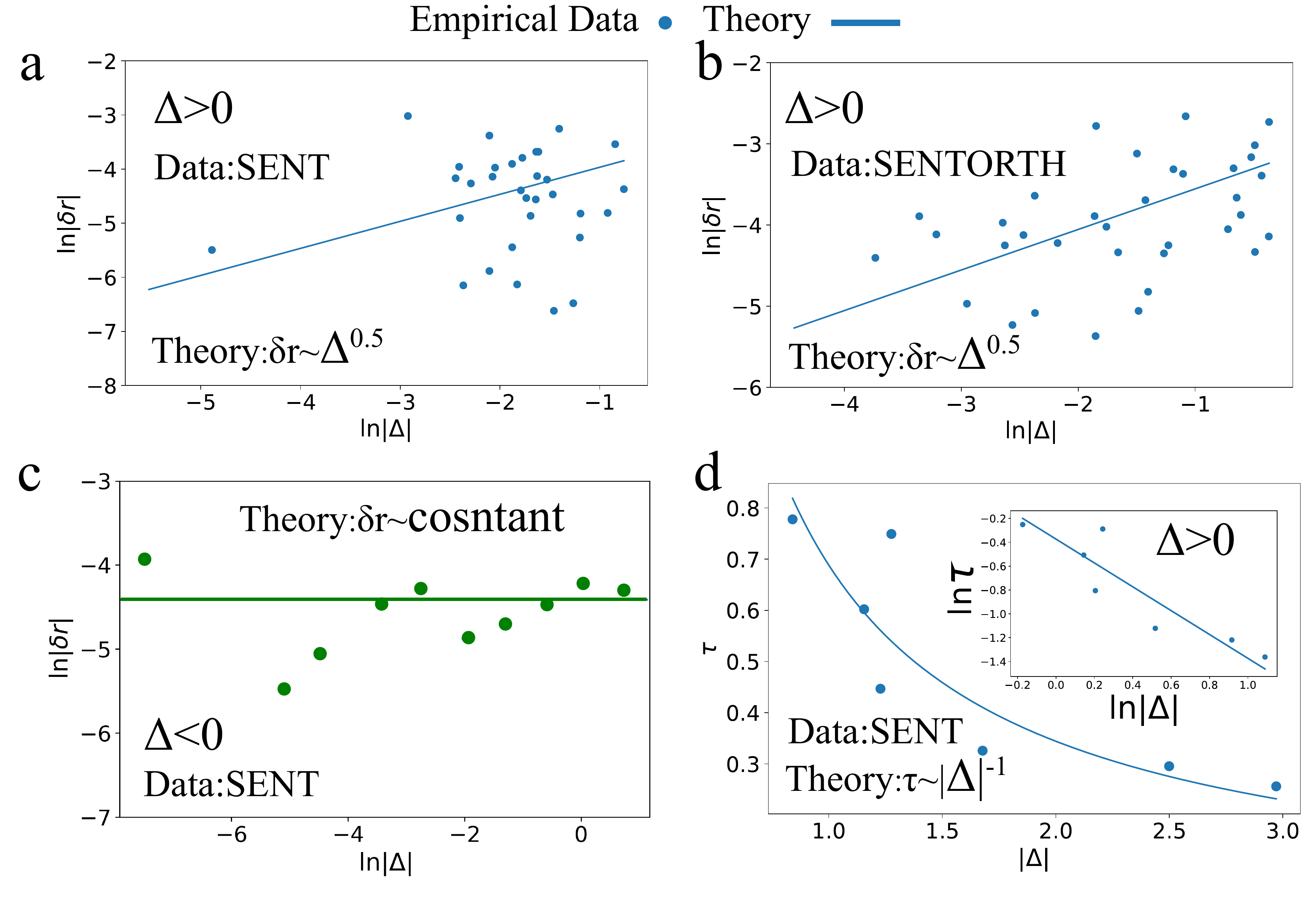}
  \caption{Market responses linked to collective sentiment through a common critical mode
(a,b) Empirical market-response magnitude $|\delta r|$ as a function of the distance from the critical threshold on the collectively biased side (blue, $\Delta>0$), using the Baker--Wurgler SENT and SENTORTH indices, respectively. The solid lines indicate the theoretical predicted square-root scaling, $|\delta r|\sim \Delta^{1/2}$.
(c) On the neutral-like side (green, $\Delta<0$), the market-response magnitude exhibits no comparable critical growth and remains approximately constant over the examined range.
(d) Empirical relaxation time $\tau$ as a function of the distance from criticality, with the inset showing the corresponding logarithmic scaling consistent with $\tau\sim |\Delta|^{-1}$.  The correspondence between the two regimes identified by $\Delta$ and the historical-event classification in Fig.~2(a) connects the collective sentiment transition to its macroscopic market signatures.
}
  \label{fig3}
\end{figure}
At the microscopic level, each investor adjusts their strategy through a combination of market-driven adaptation, intrinsic behavioral stabilization, and stochastic fluctuations. Specifically, the strategy \(w_i\in[-1,1]\) of investor \(i\) evolves according to the Langevin equation
\begin{equation}
dw_i
=(1-w^2)(a r(t)-bw)dt+\sqrt{2D}B_i
\label{eq:langevin}
\end{equation}
where \(a\) quantifies the sensitivity of individual strategies to market returns, \(b\) describes their tendency to relax toward neutrality, and \(D\) represents stochastic behavioral fluctuations. The factor \(1-w_i^2\) introduces a state-dependent responsiveness, such that both market-driven adaptation and intrinsic behavioral relaxation weaken as individual strategies approach the extreme states \(w_i=\pm1\), where the deterministic drift vanishes. Equation (\ref{eq:langevin}) leads to the heterogeneous investor population by the probability density \(\rho(w,t)\) of individual strategy biases, with collective sentiment defined as \(W(t)=\int_{-1}^{1}w\rho(w,t)\,dw\). The evolution of the strategy distribution is governed by (see Supplemental Material~\cite{LiuSupplementalMaterial}, Sec.~1.1)
\begin{equation}
\partial_t\rho
=
-\partial_w
\left[
(1-w^2)\bigl(a r(t)-bw\bigr)\rho(w,t)
\right]
+
D\,\partial_w^2\rho(w,t),
\label{eq:fokkerplanck}
\end{equation}

Meanwhile, the market return, in turn, evolves according to
\begin{equation}
  \dot r(t)=-\alpha r(t)+\beta W(t)+\sigma\xi(t),
\label{eq:return}
\end{equation}
where \(\alpha\) sets the relaxation rate of market returns, \(\beta\) characterizes their response to collective sentiment, and \(\sigma\xi(t)\) represents external market fluctuations. The reciprocal coupling between Eqs. (\ref{eq:fokkerplanck}) and (\ref{eq:return}) gives the market a dual dynamical role. It responds to the collective organization of investor strategies while simultaneously mediating the feedback through which individual strategies become collectively coupled. This structure allows collective behavioral organization to emerge without prescribing direct alignment interactions among investors and provides the basis for linking the instability of investor sentiment to macroscopic market responses.

To determine whether adaptive investor-market feedback can sustain a collective behavioral bias, we examine the stability of the neutral state. We consider the regime in which market returns relax faster than individual strategies adapt. On the slower behavioral time scale, the market response approximately follows collective sentiment, yielding \(r\simeq(\beta/\alpha)W\) from Eq. (\ref{eq:return}). This relation captures how the collective organization of investor strategies is transmitted to market returns, which in turn influence subsequent strategy adjustments. Substituting it into Eq. (\ref{eq:fokkerplanck}), projecting the dynamics onto the collective mode \(W\), and expanding near the neutral state give the leading-order effective dynamics (see Supplemental Material~\cite{LiuSupplementalMaterial}, Sec.~1.2),
\begin{equation}
   \dot W=\Delta W-uW^3+\eta_W(t),
\label{eq:W} 
\end{equation}
where \(u>0\) describes nonlinear saturation of the collective bias and \(\eta_W(t)\) represents effective stochastic forcing. The stability of the collective state is governed by
\begin{equation}
\Delta=\frac{a\beta}{\alpha}-K_c(b,D),
\qquad
K_c(b,D)=
\frac{D}{\left\langle w^2-w^4/3\right\rangle_0},
\label{eq:Delta}
\end{equation}
where $\langle\cdot\rangle_0$ denotes an average over the neutral stationary strategy distribution $\rho_0(w)\propto \exp[-(b/D (w^2/2-w^4/4)]$. The critical coupling $K_c$ follows from the exact linear response of the stationary Fokker--Planck equation (see Supplemental Material~\cite{LiuSupplementalMaterial}, Sec.~1.2).

Equation (\ref{eq:Delta}) identifies the emergence of market-mediated collective self-organization as a competition between the emergent coupling ${a\beta}/{\alpha}$ and behavioral stabilization. For \(\Delta<0\), the neutral collective state is stable and deviations of \(W\) decay; for \(\Delta>0\), feedback-driven amplification destabilizes this state, allowing persistent directional sentiment to emerge. At \(\Delta=0\), the restoring rate of the collective mode vanishes, defining the critical point of the reduced dynamics. Importantly, this mode is not confined to investor behavior. Because market returns follow collective sentiment through \(r\simeq(\beta/\alpha)W\), the same instability is transmitted to the macroscopic market response. The collective transition and its financial manifestation therefore originate from a shared dynamical mode.
\begin{figure}
  \includegraphics[width=1\linewidth]{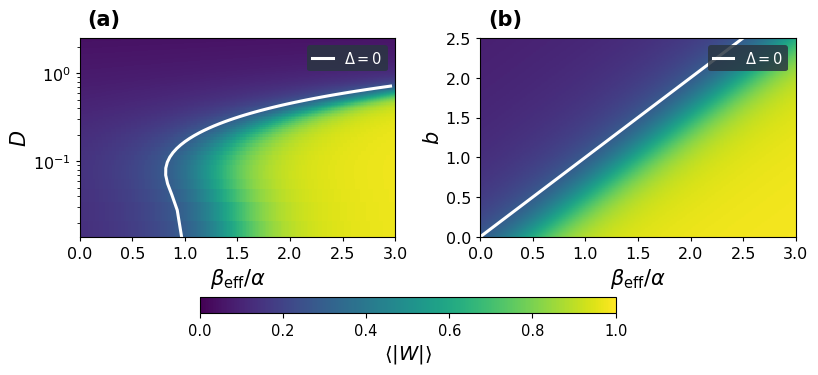}
\caption{Phase boundary of the feedback-driven collective transition.(a,b) Corresponding phase diagrams in the
$(a\beta/\alpha,D)$ and $(a\beta/\alpha,b)$
planes, respectively.
The phase boundary is determined by
$a\beta/\alpha=K_c(b,D)$.
}
  \label{fig4}
\end{figure}

Equation (4) can be written in the form of a stochastic relaxation dynamics governed by an effective potential,
\begin{equation}
\dot{W}=-\frac{\partial U(W)}{\partial W}
+\eta_W(t),
\qquad
U(W)=-\frac{\Delta}{2}W^2+\frac{u}{4}W^4.
\end{equation}
Assuming the effective noise $\eta_W$ is Gaussian and short-correlated, the corresponding stationary distribution is $P_{\rm st}(W)\propto\exp\!\left[-\frac{U(W)}{D_W}\right]$, where $D_W$ denotes the effective noise strength of the collective mode. The shape of $P_{\rm st}(W)$ is therefore controlled by $U(W)$; hence, the quantity $\Delta$ governs the stability of the neutral state.

The collective states predicted by the effective theory can be compared with the empirical sentiment distributions in Fig.~\ref{fig2}(a). During neutral-like periods, the empirical distribution retains substantial probability weight near zero, whereas the collectively biased regime exhibits a pronounced shift toward finite sentiment. As shown in Fig.~\ref{fig2}(b), the theoretical stationary distributions and microscopic simulations reproduce the corresponding reorganization from a neutral to a collectively biased state as $\Delta$ changes sign. This qualitative agreement connects the event-anchored empirical observations to the collective states generated by adaptive investor--market feedback.


To characterize the transition quantitatively, we examine the stationary states of Eq.~(\ref{eq:W}),
\begin{equation}
W_{st}\sim
\begin{cases}
0, & \Delta\leq0,\\[4pt]
\pm\sqrt{\dfrac{\Delta}{u}}, & \Delta>0,
\end{cases}
\label{eq:Wstar}
\end{equation}
revealing a continuous transition at $\Delta=0$. The square-root onset results from the competition between feedback-driven amplification and nonlinear saturation of collective behavior. As shown in Fig.~\ref{fig2}(c), microscopic simulations reproduce the predicted onset of collective sentiment, confirming that the transition persists beyond the reduced description. Together, Figures.~\ref{fig2}(a)-(c) connect the empirically observed sentiment regimes with the collective transition generated by the model.

Crucially, the collective transition and macroscopic market responses are governed by the same critical mode. In the fast-market limit, $r\simeq(\beta/\alpha)W$, the market response is directly coupled to the same collective mode responsible for sentiment organization. Consequently, the onset of
collective sentiment and the associated market response are controlled by the same critical parameter $\Delta$. The market response therefore inherits the square-root scaling of the collective order parameter (see Supplemental Material~\cite{LiuSupplementalMaterial}, Sec.~1.2),
\begin{equation}
|\delta r|\sim\Delta^{1/2},\qquad \Delta>0.
\label{eq:deltar}  
\end{equation}
This relation identifies critical market responses as macroscopic manifestations of collective behavioral reorganization. To examine this prediction empirically, we measure the market response through the difference between short- and long-term average returns, $\delta r$. Figures~\ref{fig3}(a) and (b) show behavior consistent with the predicted square-root scaling on the collectively biased side, using the SENT and SENT\_ORTH indices, respectively. In contrast, on the neutral-like side, the response remains approximately constant over the examined range (Fig.~\ref{fig3}(c)), without comparable critical growth. The empirical threshold separating these response regimes is also consistent with the event-based classification of sentiment states in Fig.~\ref{fig2}(a), linking the collective behavioral reorganization to the observed market-response scaling (see Supplemental Material~\cite{LiuSupplementalMaterial}, Sec.~2.2-2.3).

The same collective critical mode also governs the relaxation dynamics near the transition. Linearizing Eq. (3) around its stationary states shows that the restoring rate approaches zero as ($\Delta\to0$), yielding a divergent relaxation time, 
\begin{equation}
    \tau\sim|\Delta|^{-1}.
\end{equation}
where $\tau$ is the collective relaxation time.  This divergence
reflects the dynamical loss of restoring stability near the transition. Figure~\ref{fig3}(d) shows increasing empirical relaxation times as the critical threshold is approached, with the logarithmic representation consistent with the theoretical predicted inverse-distance scaling. Together, Figures~\ref{fig2} and \ref{fig3} support a unified critical interpretation: collective sentiment reorganization, macroscopic market-response scaling, and critical slowing down are linked through the same collective mode generated by adaptive investor--market feedback.

The market-mediated origin of the collective transition can be further characterized through its critical boundary in parameter space. Equation~(\ref{eq:Delta}) shows that the emergent coupling $a\beta/\alpha$ competes with a critical threshold $K_c(b,D)$ determined by the stationary response of individual strategies. The two-dimensional sections in Figs.~\ref{fig4}(a) and (b) reveal the distinct roles of behavioral restoration and stochastic fluctuations. While stronger restoration raises the critical coupling, the effect of noise is nonmonotonic. Weak fluctuations can lower the threshold for collective ordering, whereas stronger noise eventually suppresses it. This behavior arises because restoration and noise jointly reshape the stationary strategy distribution and hence its susceptibility to market feedback. The resulting phase boundaries demonstrate that collective self-organization is governed not by additive stabilizing effects, but by the interplay between market-mediated reinforcement and microscopic behavioral fluctuations.

{\it Discussion.} A central challenge in behavioral finance is to understand how individual behavioral biases become collectively amplified and translate into macroscopic market instability. The dynamical conditions under which adaptive investor-market feedback produces a collective critical transition, and how such a transition manifests in observable market responses, remain insufficiently understood. Our results address this gap by identifying a feedback-driven critical condition for the emergence of persistent collective sentiment and linking the resulting transition to quantitative signatures in financial market dynamics. This establishes a statistical-physics perspective in which market criticality can be understood as a macroscopic manifestation of the reorganization of collective investor behavior, rather than solely as a statistical property of financial fluctuations.

The key mechanism is market-mediated collective self-organization. Although direct alignment interactions among investors are not imposed, their reciprocal adaptation with market returns generates an effective collective coupling. Individual strategies shape the market response, which in turn feeds back onto the entire investor population. When this emergent reinforcement overcomes behavioral stabilization, the neutral collective state loses stability and a collective critical mode emerges. Importantly, this mode connects two distinct dynamical processes. Near the transition, the slow evolution of collective sentiment governs the onset of persistent behavioral organization and the divergence of relaxation times, while the market response inherits its critical scaling through the same investor–market coupling. Thus, the critical behavior of investor sentiment and macroscopic market returns is not governed by two independent instabilities, but is linked through a shared collective mode generated by their reciprocal dynamics.

Real financial markets involve additional mechanisms, including heterogeneous information, institutional constraints, and evolving interaction structures, which may modify the location and observable signatures of the transition. Moreover, while the empirical scaling and relaxation patterns are consistent with the predicted critical dynamics, they do not by themselves establish the proposed feedback mechanism as their unique cause. Nevertheless, the agreement between analytical predictions, microscopic simulations, and empirical observations suggests a testable route toward understanding financial market criticality. More broadly, our results show how adaptive coupling between individual decisions and aggregate market outcomes can organize investor behavior into a collective critical state, providing a dynamical link between behavioral finance and the statistical physics of nonequilibrium complex systems.

\bibliography{ref}
\end{document}


\bibliographystyle{unsrt}
\maketitle
\section{Theoretical Derivations}
\subsection{Microscopic stochastic dynamics and mean-field description}
\label{sec:sm_microscopic}

We consider a population of investors whose trading strategies
are represented by continuous variables $w_i(t)\in[-1,1]$,
where $w_i>0$ and $w_i<0$ correspond to bullish and bearish
positions, respectively, and $w_i=0$ represents a neutral
strategy. The collective investor sentiment is characterized
by the population-averaged strategy,
\begin{equation}
    W_N(t)=\frac{1}{N}\sum_{i=1}^{N}w_i(t).
    \label{eq:sm_WN}
\end{equation}

The strategy of each investor evolves under the combined
influence of market returns, individual behavioral
stabilization, and stochastic fluctuations. Consistent
with the microscopic dynamics underlying the main text,
we write
\begin{equation}
    dw_i(t)
    =(1-w_i^2)\left[a r(t)-b w_i(t)\right]dt
    +\sqrt{2D}\,dB_i(t),
    \label{eq:sm_investor_sde}
\end{equation}
where $a>0$ quantifies the sensitivity of investor
strategies to market returns, $b>0$ represents the
tendency toward neutral strategies, and $D>0$ controls
the intensity of behavioral fluctuations. The processes
$B_i(t)$ are independent standard Brownian motions.

The factor $1-w_i^2$ suppresses deterministic strategy
adjustments near the extreme bullish and bearish states.
Since the stochastic term remains nonzero at $w_i=\pm1$,
the bounded strategy domain requires an additional
boundary prescription. Here we impose reflecting
boundaries at $w_i=\pm1$, so that probability cannot
flow outside the admissible strategy interval.

The market return $r(t)$ evolves according to
\begin{equation}
    dr(t)
    =\left[-\alpha r(t)+\beta W_N(t)\right]dt
    +\sigma\,dB_r(t),
    \label{eq:sm_market_sde}
\end{equation}
where $\alpha>0$ denotes the relaxation rate of market
returns, $\beta>0$ measures the aggregate influence
of investor sentiment on market returns, and $\sigma$
sets the amplitude of market fluctuations.
The Brownian motion $B_r(t)$ is independent of the
individual behavioral noises.

Equations~(\ref{eq:sm_investor_sde}) and
(\ref{eq:sm_market_sde}) define a reciprocal
investor--market feedback loop. Individual strategies
collectively influence market returns through $W_N$,
while the resulting market return feeds back into
the evolution of every investor's strategy.
Importantly, no explicit pairwise alignment term
between investors is introduced.

In the mean-field limit $N\to\infty$, we introduce
the normalized strategy density $\rho(w,t)$,
satisfying
\begin{equation}
    \int_{-1}^{1}\rho(w,t)\,dw=1.
    \label{eq:sm_normalization}
\end{equation}
Its collective order parameter is
\begin{equation}
    W(t)=\int_{-1}^{1}w\rho(w,t)\,dw.
    \label{eq:sm_W}
\end{equation}

To derive the mean-field description, we introduce the drift function
\begin{equation}
f(w,r)=(1-w^2)(ar-bw).
\end{equation}
For a twice continuously differentiable test function $\phi(w)$ satisfying $\phi'(\pm1)=0$, It\^o's formula gives
\begin{equation}
d\phi(w_i)
=
\left[
f(w_i,r)\phi'(w_i)
+D\phi''(w_i)
\right]dt
+\sqrt{2D}\phi'(w_i)dB_i.
\end{equation}
Conditioning on the common market-return trajectory
and taking the mean-field limit $N\to\infty$, we obtain
\begin{equation}
\frac{d}{dt}\int_{-1}^{1}\phi(w)\rho(w,t)\,dw
=
\int_{-1}^{1}
\left[
f(w,r)\phi'(w)+D\phi''(w)
\right]\rho(w,t)\,dw.
\end{equation}
Integration by parts yields the Fokker--Planck equation
\begin{equation}
\partial_t\rho
=
-\partial_w\left[f(w,r)\rho\right]
+D\partial_w^2\rho,
\label{eq:sm_fp}
\end{equation}
where
\begin{equation}
    J(w,t)
    =(1-w^2)(ar(t)-bw)\rho(w,t)
    -D\frac{\partial\rho(w,t)}{\partial w},
    \label{eq:sm_current}
\end{equation}
the reflecting boundary conditions take the form
\begin{equation}
    J(-1,t)=J(1,t)=0.
    \label{eq:sm_reflecting}
\end{equation}
These conditions ensure conservation of the total probability within the bounded strategy domain. In the mean-field description, the market dynamics becomes
\begin{equation}
    \frac{dr(t)}{dt}
    =-\alpha r(t)+\beta W(t)+\sigma\xi(t),
    \label{eq:sm_market_mf}
\end{equation}
where $\xi(t)$ denotes Gaussian white noise with
$\langle\xi(t)\xi(t')\rangle=\delta(t-t')$.

The coupled equations~(\ref{eq:sm_fp}) and (\ref{eq:sm_market_mf}) constitute the starting point for the theoretical analysis. They retain the feedback between the collective investor distribution and the market return, while individual investors interact only indirectly through their common market environment.

\subsection{Collective dynamics and criticality}
\label{sec:S12}
In the fast-market limit, the return variable
adiabatically follows collective sentiment,
$r\simeq(\beta/\alpha)W$. Defining the effective
coupling $K=a\beta/\alpha$, Equation~(\ref{eq:sm_fp}) becomes
\begin{equation}
\partial_t\rho
=-\partial_w\left[
(1-w^2)(KW-bw)\rho
\right]+D\partial_w^2\rho.
\label{eq:S12_FP}
\end{equation}
Under reflecting boundary conditions, the
stationary probability current vanishes,
yielding
\begin{equation}
\rho_{\mathrm{st}}(w;W)
=\frac{1}{Z(W)}
\exp\left[
\frac{KW}{D}
\left(w-\frac{w^3}{3}\right)
-\frac{b}{D}
\left(\frac{w^2}{2}-\frac{w^4}{4}\right)
\right].
\label{eq:S12_rho}
\end{equation}
The stationary collective sentiment satisfies
the self-consistency condition
\begin{equation}
W=\int_{-1}^{1}
w\rho_{\mathrm{st}}(w;W)\,dw.
\label{eq:S12_self}
\end{equation}
Linearizing this condition around the
symmetric state $W=0$ gives
\begin{equation}
1=\frac{K_c}{D}
\left\langle w^2-\frac{w^4}{3}
\right\rangle_0,
\label{eq:S12_critical}
\end{equation}
where $\langle\cdot\rangle_0$ denotes an
average over the neutral stationary density
\begin{equation}
\rho_0(w)\propto
\exp\left[
-\frac{b}{D}
\left(\frac{w^2}{2}-\frac{w^4}{4}\right)
\right].
\end{equation}
The critical coupling is therefore
\begin{equation}
K_c(b,D)=
\frac{D}{
\left\langle w^2-w^4/3\right\rangle_0
},
\qquad
\Delta=K-K_c(b,D).
\label{eq:S12_Delta}
\end{equation}

Near this critical boundary, the collective
dynamics is dominated by the slowly relaxing
symmetry-breaking mode. Because the microscopic
equations are invariant under
$(w,W,r)\rightarrow(-w,-W,-r)$, the reduced
dynamics is odd in $W$. Retaining the leading
linear and nonlinear terms gives, after
normalizing the time scale,
\begin{equation}
\frac{dW}{dt}
=\Delta W-uW^3+\eta_W(t),
\label{eq:S12_Landau}
\end{equation}
where $u>0$ represents nonlinear saturation
in the supercritical regime and $\eta_W(t)$
denotes effective stochastic forcing.

Neglecting the stochastic forcing, the
stationary states satisfy
\begin{equation}
0=W(\Delta-uW^2).
\end{equation}
Their stable branches are
\begin{equation}
W_{\mathrm{st}}=
\begin{cases}
0, & \Delta\leq0,\\[2pt]
\pm\sqrt{\Delta/u}, & \Delta>0.
\end{cases}
\label{eq:S12_Wst}
\end{equation}
Thus, the collective order parameter
emerges continuously with the mean-field
exponent $1/2$.
Since the market return adiabatically follows
collective sentiment, $r\simeq(\beta/\alpha)W$,
the stationary market response satisfies
\begin{equation}
|\delta r|
\simeq\frac{\beta}{\alpha}|W_{\mathrm{st}}|
=\frac{\beta}{\alpha}\sqrt{\frac{\Delta}{u}}
\sim\Delta^{1/2},
\qquad \Delta>0.
\end{equation}
Thus, the collective ordering transition
directly induces the same critical exponent
in the macroscopic market response.

Finally, linearizing the reduced dynamics
around a stable stationary state gives
\begin{equation}
\frac{d\,\delta W}{dt}
=\left(\Delta-3uW_{\mathrm{st}}^2\right)
\delta W.
\end{equation}
The corresponding relaxation time obeys
\begin{equation}
\tau\sim|\Delta|^{-1},
\label{eq:S12_tau}
\end{equation}
demonstrating critical slowing down as
the collective transition is approached.

\section{Empirical data description}
\subsection{Data sources and sample coverage}
\label{sec:S21}

Our empirical analysis combines U.S. stock-market
returns from the Kenneth R. French Data Library
with the Baker--Wurgler investor sentiment indices
\cite{fama1993common,baker2006investor}. The market data are drawn
from the Fama--French Research Data Factors,
available at daily, weekly, and monthly frequencies.
We use the market excess return (Mkt-RF) and
the risk-free rate (RF), both reported in percent.
For the market-return analysis, the total market
return is constructed as
$r_t^{\mathrm{mkt}}=(\mathrm{Mkt\!-\!RF})_t+
\mathrm{RF}_t$, with returns converted to
decimal units. The monthly excess-return series
is also used in the event-based analysis.

Investor sentiment is measured using the monthly
Baker--Wurgler indices. We consider both the
original sentiment index (SENT) and its
macroeconomically orthogonalized counterpart
(SENT-ORTH). The updated indices are constructed
from five market-based sentiment proxies, rather
than direct observations of individual trading
strategies. We therefore interpret them as
empirical proxies for aggregate investor sentiment.
The available sentiment-index sample extends
from July 1965 through December 2023, defining
the common monthly coverage with market returns.
For analyses involving daily or weekly returns,
each observation is matched to the sentiment
index of its corresponding calendar month.
The data sources and coverage are summarized
in Table~\ref{tab:empirical_data}.

\begin{table}[t]
\caption{Summary of empirical datasets.}
\label{tab:empirical_data}
\centering
\begin{tabular}{llll}
Dataset & Frequency & Sample period & Variables \\
\hline
Fama--French factors & Monthly &
1926/07--2026/01 & Mkt-RF, RF \\
Fama--French factors & Weekly &
1926/07--2026/01 & Mkt-RF, RF \\
Fama--French factors & Daily &
1926/07--2026/01 & Mkt-RF, RF \\
Baker--Wurgler sentiment & Monthly &
1965/07--2023/12 & SENT, SENT-ORTH \\
\end{tabular}
\end{table}

\begin{table}[t]

\caption{
Historical event anchors used to construct
the empirical sentiment distributions. The historical
classifications are based on the cited
sources.
}
\label{tab:S1_events}
\centering
\begin{tabular}{p{2.8cm}p{7.0cm}p{3.0cm}}
\hline\hline
Anchor months & Historical episode & References \\
\hline
1969-12
& U.S. recession of 1969--1970
& NBER \cite{NBERCycles} \\

1981-11, 1981-12, 1982-01
& U.S. recession of 1981--1982
& NBER \cite{NBERCycles} \\

1985-01
& Strong U.S. equity-market rally and rising investor optimism in January 1985
& \cite{Ross1985,LATimes1985} \\

1986-09, 1986-10,
1987-02, 1987-03,
1987-04
& Market expansion preceding the
1987 stock-market crash
& Federal Reserve \cite{Fed1987} \\

2000-03, 2000-06
& Dot-com bubble collapse
& Lansing \cite{Lansing2003} \\

2001-12
& Post-dot-com market adjustment
& Bernanke \cite{Bernanke2010} \\

2008-03
& Bear Stearns liquidity crisis
& Federal Reserve \cite{FedBear2008} \\

2021-02
& Retail-investor and meme-stock
trading episode
& SEC \cite{SEC2021} \\

2022-06, 2022-07,
2022-08
& Inflation, monetary tightening,
and equity-market adjustment
& Cotton \cite{Cotton2022} \\
\hline\hline
\end{tabular}
\end{table}

\subsection{Construction of empirical observables}
\label{sec:S22}

To establish the empirical connection between collective sentiment organization and market criticality, we first examine
the distribution of investor sentiment around historically documented episodes of market stress and speculative activity. The event anchors, summarized in Table~\ref{tab:S1_events}, cover major
market disruptions and periods of pronounced speculative behavior, including the dot-com bubble, the global financial crisis, and the
2021 retail-investor trading episode. For each event anchor, we extract monthly Baker--Wurgler sentiment observations within a symmetric six-month window. To compare the empirical sentiment distributions with the theoretical bounded strategy variable, the sentiment index is linearly rescaled to $[-1,1]$ using its full-sample minimum and maximum. Kernel density estimates are obtained separately for each window and then averaged within the event and reference groups. These procedures yield the collectively biased and neutral-like sentiment distributions shown in Fig~2(a).

The critical-response analysis uses the standardized sentiment index
\begin{equation}
x_t=\frac{x_t^{\mathrm{raw}}-\mu_x}{s_x},
\label{eq:S22_sentiment}
\end{equation}
where the mean $\mu_x$ and standard deviation $s_x$ are calculated separately for SENT and SENT\_ORTH. The standardized variable $x_t$ provides a common empirical coordinate for locating the collective sentiment regimes and examining their associated market responses.

We denote the empirical critical threshold by $x_c$ and define the signed distance
\begin{equation}
d_t=x_t-x_c.
\label{eq:S22_distance}
\end{equation}
Importantly, the resulting division between $d_t<0$ and $d_t>0$ is consistent with the neutral-like and collectively biased sentiment regimes independently identified from the historical event analysis in Fig.~2(a). 

The second observable measures deviations of
market returns from their longer-term behavior.
Let $r_t^{\mathrm{mkt}}$ denote the monthly total
market return, constructed by adding the market
excess return to the risk-free rate. We define
\begin{equation}
\delta r_t
=
\overline{r}^{(3)}_t
-
\overline{r}^{(24)}_t,
\label{eq:S22_delta_r}
\end{equation}
where $\overline{r}^{(m)}_t$ is the rolling average
of monthly market returns over the most recent
$m$ months. The response amplitude used to test
the predicted square-root scaling is
$|\delta r_t|$.

To characterize the dynamical response, we further
define the market-return deviation from a longer-term
rolling baseline,
\begin{equation}
u_t=r_t^{\mathrm{mkt}}-
\overline{r}^{(L)}_t,
\label{eq:S22_u}
\end{equation}
where $L$ is the baseline window used in the
relaxation-time analysis. We estimate the
persistence of $u_t$ using an autoregressive
model of order one,
\begin{equation}
u_t=c+\phi u_{t-1}+\epsilon_t,
\label{eq:S22_AR1}
\end{equation}
where $c$ is an intercept and $\epsilon_t$ is
a residual. For $0<\phi<1$, the corresponding
relaxation time is defined as
\begin{equation}
\tau=-\frac{1}{\ln\phi},
\label{eq:S22_tau}
\end{equation}
measured in months. This estimate quantifies
the characteristic decay time of the return
deviation within the corresponding empirical
subsample.

\subsection{Robustness of critical market responses and slowing down}
\newpage

\begin{figure}
  \includegraphics[width=1\linewidth]{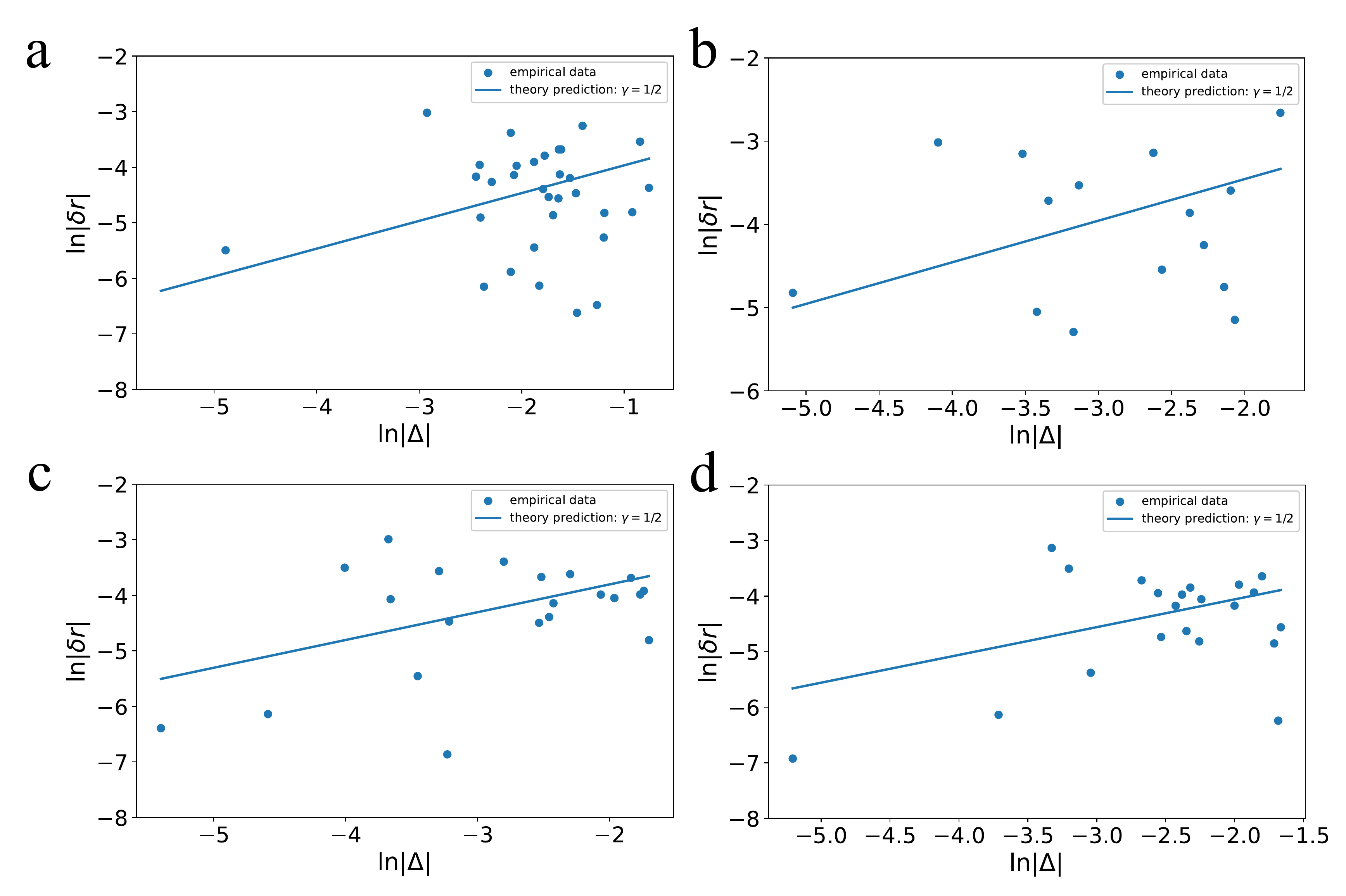}
\caption{
\textbf{Robustness of critical market-response scaling
under alternative return-averaging windows (SENT).}
Logarithmic market-response magnitude
$\ln|\delta r|$ versus the logarithmic distance
from the empirical critical threshold,
$\ln|d|$, using the Baker--Wurgler SENT index.
The market response is defined as the difference
between short- and long-term moving-average returns,
$\delta r=\bar r^{(m_s)}-\bar r^{(m_l)}$.
The averaging windows $(m_l,m_s)$, in months,
are (a) $(24,3)$, (b) $(24,6)$,
(c) $(36,3)$, and (d) $(36,6)$.
Solid lines show the theoretical square-root
scaling with exponent $1/2$.
The results exhibit broadly consistent positive
scaling trends across alternative averaging windows.
}
\end{figure}
\clearpage
\begin{figure}
  \includegraphics[width=1\linewidth]{RobustnessCheckDeltaR(1).pdf}
\caption{
\textbf{Robustness of critical market-response scaling
under alternative return-averaging windows
(SENT\_ORTH).}, using the
orthogonalized Baker--Wurgler sentiment index
SENT\_ORTH.
The long- and short-term averaging windows
$(m_l,m_s)$, in months, are
(a) $(24,3)$, (b) $(24,6)$,
(c) $(36,3)$, and (d) $(36,6)$.
Solid lines represent the predicted
square-root scaling with exponent $1/2$.
The positive scaling trend is retained
under this alternative sentiment measure,
providing an additional check on the
dependence of the results on the sentiment index.
}
\end{figure}
\clearpage

\begin{figure}
  \includegraphics[width=1\linewidth]{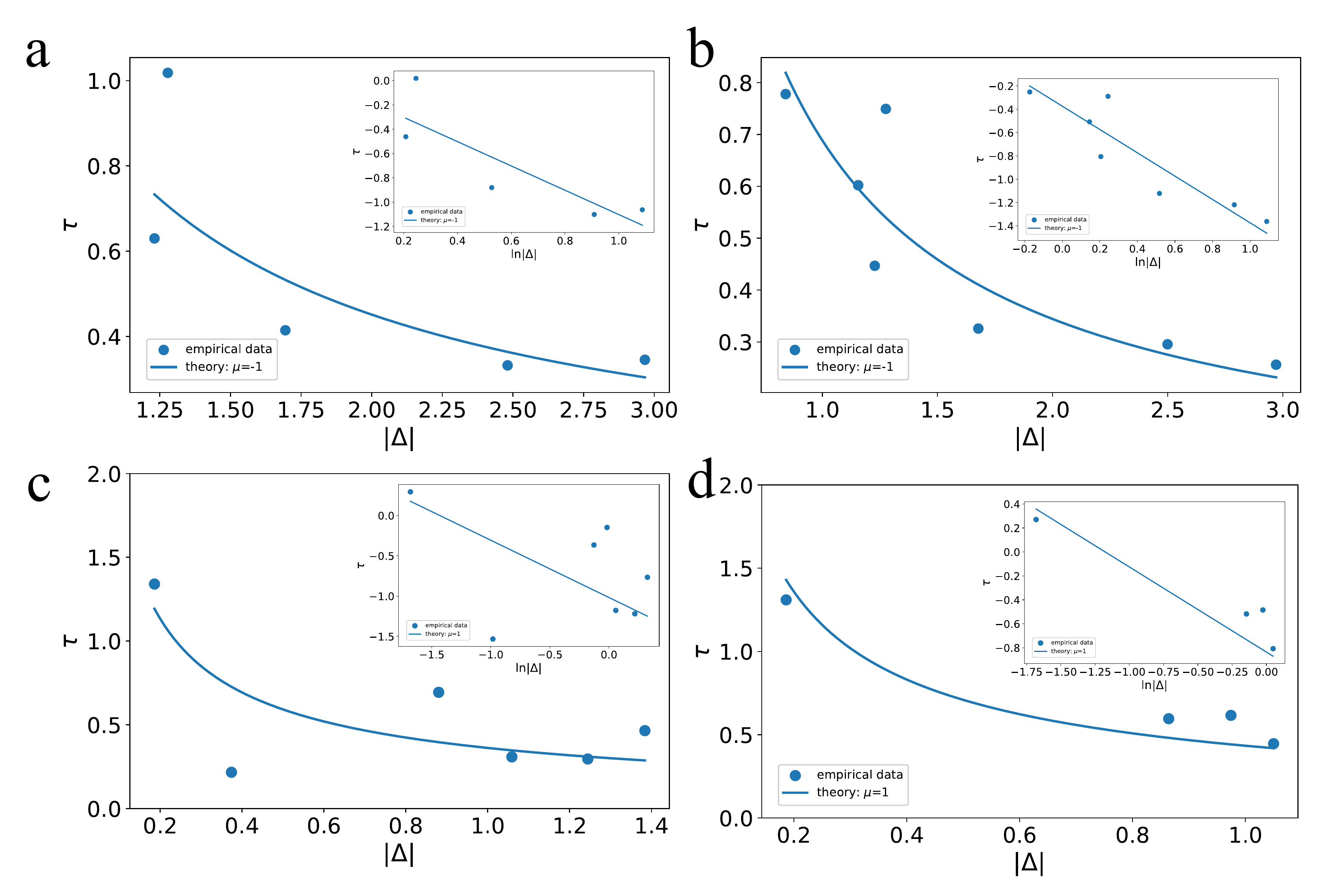}
\caption{
\textbf{Robustness of critical slowing down
across sentiment measures and detrending windows.}
Empirical relaxation time $\tau$ as a function
of the distance $|\Delta|$ from the critical
threshold, using different sentiment indices
and long-term moving-average windows for
return detrending.
The specifications are
(a) SENT with a 24-month window,
(b) SENT with a 36-month window,
(c) SENT\_ORTH with a 24-month window,
and (d) SENT\_ORTH with a 36-month window.
Solid curves show the theoretical
inverse-distance dependence
$\tau\sim|\Delta|^{-1}$.
Insets display the corresponding
logarithmic representations.
The results show a broadly decreasing
relaxation time with increasing distance
from criticality across the four specifications.
}
\end{figure}
\clearpage

\bibliography{ref}